\documentclass[12pt,a4paper,twocolumn]{narms2}

\usepackage{subfigure} 
\usepackage{psfig}
\usepackage{timesmt}   
\usepackage{graphicx}
\usepackage{amsmath}
\usepackage{amssymb}
\usepackage{chikako}   
\makeatletter
\renewcommand{\section}{\@startsection%
{section}%
{1}%
{0mm}%
{- \baselineskip}%
{0.15\baselineskip}%
{\normalfont\normalsize}}%

\renewcommand{\subsection}{\@startsection
{subsection}%
{2}%
{0mm}%
{-\baselineskip}%
{0.15\baselineskip}%
{\normalfont\normalsize}}%
\makeatother

\newcommand{\md}{{\rm d}}

\begin{document}
\title{Rolling friction and bistability of rolling motion}
\author{\large {T.~P\"oschel,~ T.~Schwager}\\
{\em Institut f\"ur Biochemie, Charit\'e, Monbijoustra{\ss}e 2, 
  D-10117 Berlin, Germany}\\[0.3cm]
\large {N.~V.~Brilliantov}\\
{\em Moscow State University, Physics Department, Moscow 119899, 
  Russia}\\[0.3cm]
\large{A.~Zaikin}\\
{\em Institut f\"ur Physik, Universit\"at Potsdam, Am Neuen Palais 10, 14469 Potsdam, Germany} 
}
\date{\vspace*{-7.9cm} {\begin{flushright}\footnotesize In: Proceedings Powders \& Grains'05, Balkema (Rotterdam, 2005)\vspace*{4cm}\end{flushright}}}

\abstract{
  The rolling motion of a rigid cylinder on an inclined flat viscous surface is investigated and the nonlinear resistance force against rolling, $F_R(v)$, is derived. For small velocities $F_R(v)$ increases with velocity due to increasing deformation rate of the surface material. For larger velocity it decreases with velocity due to decreasing contact area between the rolling cylinder and the deformed surface. The cylinder is, moreover, subjected to a viscous drag force and stochastic fluctuations due to a surrounding medium (air). For this system, in a wide range of parameters we observe bistability of the rolling motion. Depending on the material parameters, increasing the noise level may lead to increasing or decreasing average velocity.}

\maketitle
\frenchspacing   


\section{INTRODUCTION}

Rolling friction belongs to the most important dissipation mechanisms which may sensitively affect the dynamical behavior of granular systems. Since 1785 when Vince described systematic experiments to determine the nature of friction laws \nocite{Vince:1785}, rolling friction has been investigated by many scientists due to its great importance in engineering and natural sciences, e.g. \shortcite{many1,many2,many3,many4,many5,many6}.

Experimentally it was observed that the rolling friction coefficient depends non-monotonically on the velocity
\shortcite{NonlinFrict1,NonlinFrict2,NonlinFrict3,NonlinFrict4,NonlinFrict5,NonlinFrict6,NonlinFrict7}: For small velocities the rolling friction force increases with velocity, while for fast motion it decays.
In the case of a soft cylinder rolling on a hard plane \shortcite{BrilliantovPoeschelEPL}, the contact surface between the bodies is flat which allows for the application of Hertz' contact theory extended to the contact of viscoelastic particles \shortcite{BSHP}. In the opposite case, this assumption is not justified since the plane shape follows the shape of the rolling body in the contact area.
Here we investigate rolling of a hard cylinder on a soft plane and bistability of rolling motion as a consequence of the resulting friction law.


\section{MODEL}

\begin{figure}[htbp]
  \begin{center}
    \includegraphics[width=6cm,angle=0]{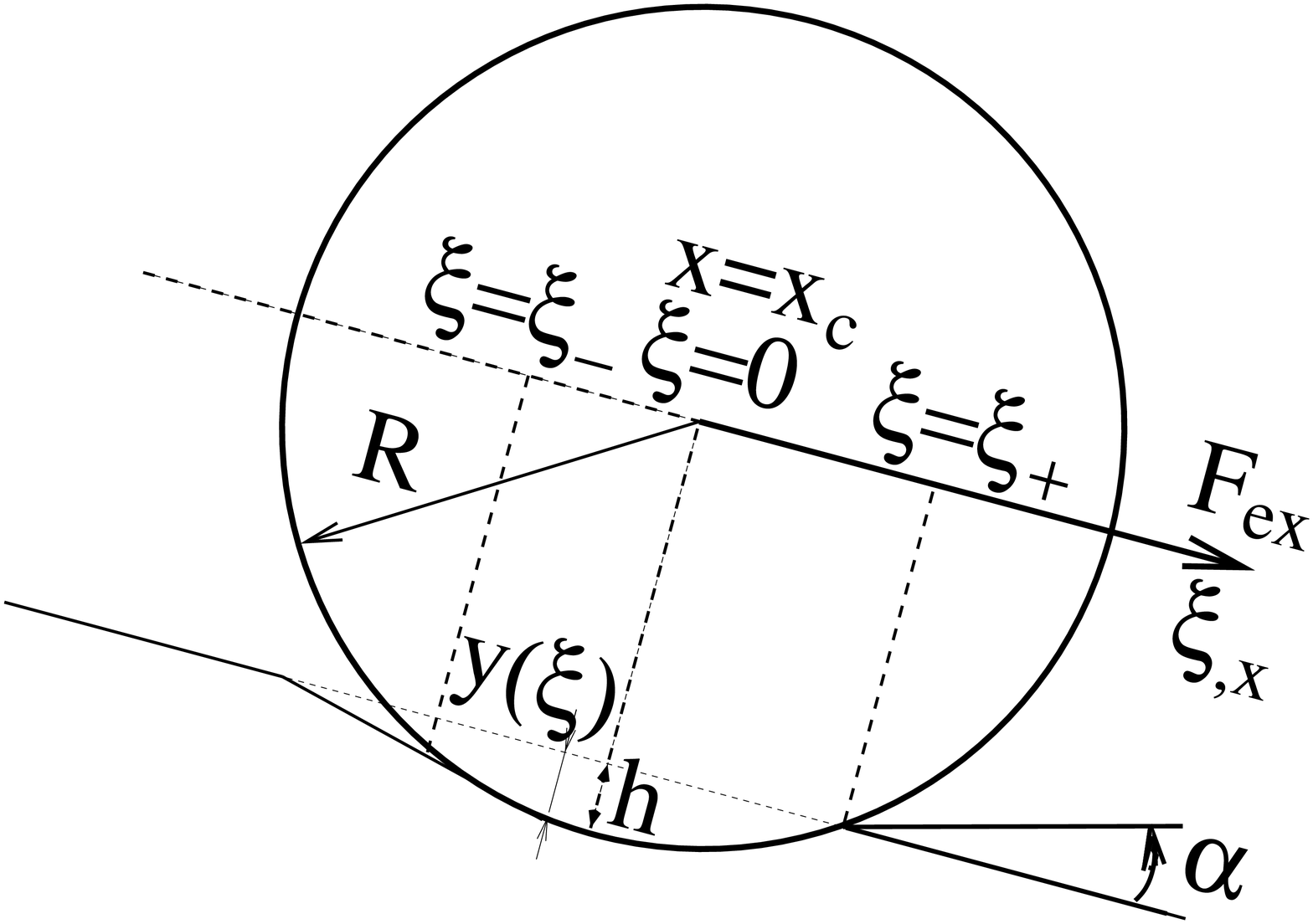}
  \end{center}
  \caption{A rigid cylinder rolls on an inclined plane built of uncoupled damped springs. The cylinder contacts the surface of the deformed plane in the region $(\xi_-,\xi_+)$ which moves along with the cylinder.}
  \label{fig1}
\end{figure}

Consider a cylinder of radius $R$, mass $M$ and moment of inertia $I$ which rolls at velocity $v$ on an inclined plane (quantities such as $M$, $I$, as well as the forces $F_R$, $F_{\rm ex}$ and similar quantities, are given as {\em per unit length} of the cylinder, i.e. as line densities), see Fig. \ref{fig1}.
For certain materials it has been shown that surface effects such as adhesion may have significant influence on rolling friction (e.g. \shortcite{adhesion1,adhesion2,adhesion3,adhesion4,adhesion5}).
On the other hand, for viscoelastic materials it was reported that rolling friction is due very little to surface interactions, i.e., the major part is due to deformation losses within the bulk of the material \shortcite{noadhesion1,noadhesion2,noadhesion3,noadhesion4}. We will, therefore, neglect surface effects. 
The surface is modeled by non-interacting springs (for the justification of this approximation see \shortcite{PSB}). The properties of the plane are described by the coefficients $k$, $\gamma$, and $m$: $k \md \xi$ and $\gamma\md \xi$ are the elastic and dissipative forces and $m\md \xi$ is the mass of the springs in the region $\md\xi$. For $x_-\le x\le x_+$ the surface is deformed by the cylinder. Assuming Hooke's law, the dynamics of the surface is described by
\begin{equation}
  \label{eq:eom}
  m\ddot{y}\left(x\right)+\gamma\dot{y}\left(x\right)+ky\left(x\right)=f\left(x,t\right)\,,
\end{equation}
where $f\left(x,t\right)$ is the density of the force which acts onto the springs in the region of contact. This force has to be provided by the cylinder. In a wide range of material parameters, the condition $k<\gamma^2/4k^2$ assures that the contact area is continuous, i.e., the cylinder contacts the surface everywhere in the interval $\left(x_-,x_+\right)$, see \shortcite{PSB} for details. 

From geometry we find the deformation $y\left(x\right)$ in the contact area $\left(x_-,x_+\right)$: 
\begin{equation}
   y\left(x\right)=R-h-\sqrt{R^2-\left(x-x_c\right)^2}\approx  \frac{\left(x-x_c\right)^2}{2R}-h\,,
\label{zxc}
\end{equation}
where $x_c(t)$ is the position of the cylinder, $h\equiv -y_{\rm min}=-y\left(x_c\right)$ is the penetration depth of the cylinder and $R\gg \left|x-x_c\right|$ was assumed. Thus, in the stationary state, $x_c=vt$, we obtain
\begin{equation}
  \dot{y}\left(x\right)=-v\frac{x-x_c}{R}\,,~~~~~~~~\ddot{y}(x)=\frac{v^2}{R}
\end{equation}
and, therefore,
\begin{equation}
\label{eqf}
f\left(\xi\right)=\frac{k}{2R}\xi^2-\frac{\gamma v}{R}\xi+\frac{mv^2}{R}-hk\,.
\end{equation}
with $\xi\equiv x-x_c$. The boundary of the contact area at the front side of the cylinder in the direction of motion follows from geometry (in the co-moving frame), 
\begin{equation}
  \label{xiplus}
\xi_+=\sqrt{2Rh} 
\end{equation}
while the other boundary follows from the contact condition $f\left(\xi_-\right)=0$:
\begin{equation}
  \label{ximinus}
  \xi_-=\frac{\gamma v}{k}-\sqrt{2hR+\left(\frac{\gamma^2}{k^2}-2\frac{m}{k}\right)v^2}\,.
\end{equation}

The springs at $x=x_+$ are special: at time $t-\delta$ ($\delta\to +0$) their velocity is zero, while an infinitesimal time later it is finite. This singularity may be attributed to a {\em finite} force $F^\prime_N$ acting at the point  $x_+$. During the time $\md t$ the cylinder moves by $v\md t$ and accelerates springs of mass $mv\md t$, 
therefore
\begin{equation}
  \label{Fspr}
  F^\prime_N=-\frac{\md p}{\md t}=-\dot{y}\left(x_+\right)m v 
=\sqrt{\frac{2h}{R}}mv^2\,.
\end{equation}
The dissipation of energy due to the plane deformation and the instantaneous acceleration of the plane material at $\xi_+$ in the instant of first contact reads 
\begin{equation}
  \dot{E}=-\int_{\xi_-}^{\xi_+}\md \xi f\left(\xi\right) \dot{y}\left(\xi\right) - \frac{m}{2}\dot{y}^2\left(\xi_+\right)v \equiv -vF_R\,,
\end{equation}
defining the force $F_R$ which acts against rolling 
\begin{equation}
  F_R=-\frac1R \int_{\xi_-}^{\xi_+}\md \xi \xi f\left(\xi\right) + m v^2\frac{h}{R}\,.
\label{FR}
\end{equation}
To evaluate $F_R$ we have to determine the penetration depth $h(v)$. It is given implicitly by equilibrating the total force exerted by the plane to the cylinder and the weight of the cylinder,
\begin{equation}
  -\int_{\xi_-}^{\xi_+}f\left(\xi\right)\md \xi +F^\prime_N=Mg\cos\alpha\,,
\label{eqh}
\end{equation}
where $f\left(\xi\right)$ and $F^\prime_N$ are given by Eqs. \eqref{eqf} and \eqref{Fspr}. 
For an analytic evaluation of $F_R$ we assume $\xi_-=0$. This approximation means that the surface recovers so slowly, that the recovering surface does not transmit energy back to the cylinder\shortcite{GreenwoodMinshallTabor:1961}. Since for realistic parameters (see below) the effects of interest, namely bistability of the rolling velocity and noise controlled velocity, occur at rather large velocity, this approximation is justified. Then Eq. \eqref{eqh} reduces to $Mg \cos \alpha =\gamma v h +\frac23 \sqrt{2R} k h^{3/2}$, which, for large damping, $v\gamma\gg k \sqrt{2Rh}$, yields
\begin{equation}
  \label{happrox}
  h(v) \approx
\frac{Mg\cos \alpha}{\gamma v+\frac23 k \sqrt{\frac{2RMg \cos \alpha}{ \gamma v}}} \, .
\end{equation}
With the same arguments from Eq. \eqref{FR} we obtain the rolling friction force
\begin{equation}
  \label{eq:Fr_an}
F_R=\frac{kh^2}{2}+\frac{mv^2h}{R}+\frac{2 \gamma v h}{3 R} \sqrt{2Rh} \, .
\end{equation}
\begin{figure}[h!]
\begin{center}
\includegraphics[width=0.35\textwidth,clip]{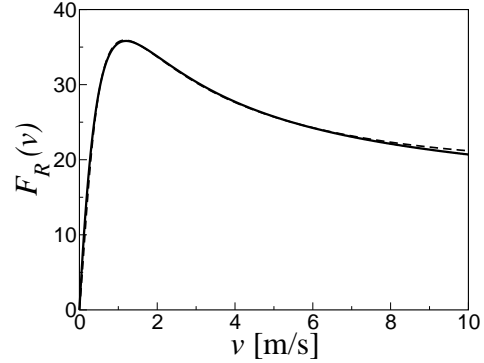}
\end{center}
\caption{
  Rolling friction force $F_R(v)$. Full line: numerical solution, dashed line: approximative theory, Eqs. \eqref{happrox} and \eqref{eq:Fr_an}.}
\label{fig2}
\end{figure}
Figure \ref{fig2} 
shows $F_R(v)$ as given by Eq. \eqref{eq:Fr_an} together with the numerical solution of Eqs. (\ref{xiplus},\ref{ximinus},\ref{FR},\ref{eqh}) for realistic material parameters, corresponding to soft rubber ($m=$100\,kg/m$^2$, $k=10^7$\,kg/m$^2$/s$^2$ , $\gamma= 5\times 10^5$\,kg/m$^2$/s). 
The faster the cylinder moves, the more efficiently energy is dissipated due to increasing deformation rate. On the other hand, the faster it moves the smaller is the contact area, i.e. the amount of deformed material decreases. Consequently, $F_R$ depends non-monotonically on the velocity, which agrees with experimental results \shortcite{NonlinFrict1,NonlinFrict2,NonlinFrict3,NonlinFrict4,NonlinFrict5,NonlinFrict6,NonlinFrict7}.
\medskip

\section{CYLINDER ROLLING DOWN AN INCLINE}

\begin{figure}[h!]
\begin{center}
\vspace*{0.2cm}
\includegraphics[width=0.35\textwidth,clip]{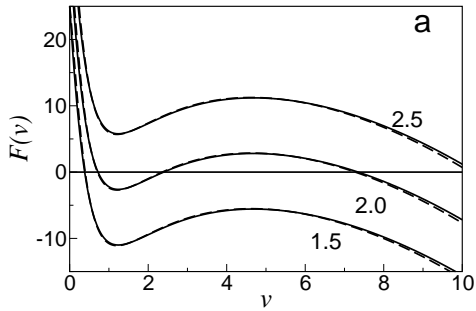}
\end{center}
\caption{
  Total force $F(v)$ for different angle of inclination, 1.5$^o$, 2$^o$, and 2.5$^o$.
Full lines: numerical solution, dashed lines:  approximative theory.}
\label{figS}
\end{figure}

The rolling friction force $F_R$ is now applied to describe the motion of a cylinder rolling down a plane inclined by the angle $\alpha$. The cylinder is subjected to an external driving force $F_{\rm ex}=Mg\sin\alpha$ ($g$ is the acceleration of gravity). The rolling friction force $F_R$ and the viscous drag force $F_D$ due to the surrounding air counteract this motion. We also assume that the tangential force acting between the cylinder and the surface at the contact area is strong enough to keep it from sliding. Newton's  equation for the cylinder reads
\begin{equation}
\label{Newton} \left(M +I/R\right) \dot{v} = -F_D(v)-F_R(v)+F_{\rm ex} +\zeta(t) \, .
\end{equation}
The viscous drag force is given by
\begin{equation}
  \label{air}
  F_D=Av + Bv^2\,.
\end{equation}
Detailed analysis \shortcite{PBZ} suggests $A=0$ and $B=0.2$\,kg/m$^2$.
Figure \ref{figS} shows the total force acting on the cylinder, ${F}(v)=-F_D-F_R+F_{\rm ex}$.

\begin{figure}[b]
\begin{center}
\includegraphics[width=0.35\textwidth,clip]{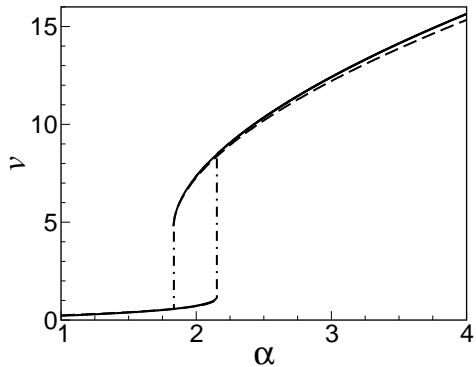}
\end{center}
\caption{
  Bifurcation diagram for stable velocity states (full line: numerical solution, dashed line: analytical approximation). 
}
\label{fig3}
\end{figure}

The steady state condition, $\dot{v}=0$, may be fulfilled either for only one velocity (top and bottom curves in Fig. \ref{figS}) or for three different velocities (middle curves). The former case implies a unique stationary velocity, while the latter allows for three stationary velocities. Only two of them, the smallest and the
largest, correspond to stable motion ($ \partial {F} /\partial v<0$). Modifying the inclination $\alpha$, the system may transit from one stable solution to the other. Figure \ref{fig3} shows the bifurcation diagram.

As shown in Figs. \ref{fig2}-\ref{fig3} the analytical theory (dashed curves) agrees well with the numerical results if the damping parameter $\gamma$ is large. 
\medskip

\section{NOISE-INDUCED JUMPS}

\begin{figure}[htbp]
\vspace*{0.2cm}
\centerline{\includegraphics[width=0.48\textwidth,clip]{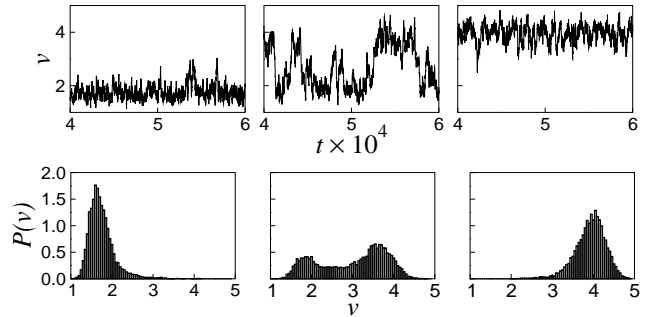}}
\caption{Top: velocity of the cylinder subjected to noise ($\sigma^2$=50\,kg$^2$/s$^3$). Bottom: corresponding velocity distributions. From left to
right: $\alpha\!=\!4.87^o$\!, $4.9^o$\!, $4.93^o$\!.  The parameters are $\gamma=0.8\times 10^5$\,kg/m$^2$/s, $k=10^6$\,kg/m$^2$/s$^2$, $B=0.65$\,kg/m$^2$.} 
\label{model2}
\end{figure}

Up to now, we did not consider noise, which is always present in realistic systems. The stochastic force  $\zeta(t)$ describing fluctuations in the media is modeled  by Gaussian white noise of zero average, $\langle \zeta(t)\rangle=0$, and intensity $\sigma$: $\langle \zeta(t)\zeta(t^\prime) \rangle=\sigma^2 \delta(t-t^\prime)$. If the system has only one stable velocity, the addition of noise does not change the motion of the cylinder qualitatively. In this case the velocity fluctuates around the average value given by the steady state condition $\dot{v}=0$. Figure \ref{model2} (top row) shows the velocity of the cylinder as obtained from numerical integration of the stochastic equation \eqref{Newton}. The corresponding velocity distribution reveals a single peak (Fig. \ref{model2}, bottom row).

For parameters corresponding to bistable velocity, the presence of noise changes the system qualitatively due to stochastic jumps between the meta-stable velocities, Fig. \ref{model2} (middle column). Consequently, the velocity distribution has two well separated peaks.

\begin{figure}[t!]
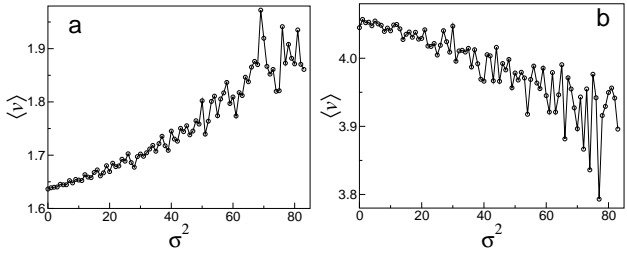

\begin{center}
\includegraphics[width=0.23\textwidth,clip]{figs/FIG7a.eps}
\includegraphics[width=0.23\textwidth,clip]{figs/FIG7b.eps}
\end{center}
\caption{
  Mean velocity of the cylinder over the noise intensity. (a)
  $\alpha=4.87^o$, (b) $\alpha=4.93^o$.}
\label{figAccel}
\end{figure}

Due to the nonlinear dependence of rolling friction on the velocity, the average velocity may increase or decrease with increasing noise level, Fig. \ref{figAccel}.
This phenomenon may be understood by analyzing the potential $U(v)\equiv -{\rm d}F/{\rm d}v$. Depending on the parameters, $U(v)$ has a double- or single-well shape \shortcite{PBZ}, corresponding to the stationary state(s) of the velocity, see Fig. \ref{figS}. In both cases, $U(v)$ is an asymmetric function in the surrounding of its minima where ${F}(v) = 0$. In the absence of noise the velocity remains in its (meta-)stable state. Subjected to noise, however, the velocity of the cylinder fluctuates around this minimum. If $U(v)$ close to the minimum is steeper in the direction of lower velocities than in the direction of higher velocities, the average velocity will be shifted towards higher velocities and, thus, increases  with increasing noise level (Fig. \ref{figAccel}a). In the opposite case, large noise impedes rolling (Fig. \ref{figAccel}b).

\section{SUMMARY}

We studied the rolling motion of a hard cylinder on an inclined viscoelastic plane in the presence of a surrounding medium. For certain realistic parameters the stationary velocity of the cylinder is bistable. For large damping of the plane's viscous deformation the numerical results agree well with theory.

In the presence of noise as is unavoidable in any realistic experiment, by means of numerical simulations we found noise-induced transitions between the meta-stable velocities. Depending on the system parameters, increasing noise level may accelerate or decelerate the rolling motion.

The described effects may be important for technical systems where the presence of noise may lead to an effective increase of the mobility of a rolling body which is driven by an external force.


\end{document}